\documentstyle[epsf]{l-aa} 

\begin{document}

\newcommand{\mic}{\,{\rm \mu m} } 

\thesaurus{08(09.04.1,     
           13.09.3,     
           09.07.1)}     

\title{Evidence for dust emission in the Warm Ionised Medium using WHAM data}

\author{G. Lagache  
	\inst{1} \and
	L.M. Haffner 
        \inst{2}  \and
        R.J. Reynolds
        \inst{2} \and
        S.L. Tufte 
        \inst{3}}

\institute{$^1$ Institut d'Astrophysique Spatial, B\^at.  121, 
Universit\'e Paris XI, 91405 Orsay Cedex, France\\
$^2$ Astronomy Department, University of Wisconsin, Madison, WI 53706\\
$^3$ Department of Physics, Lewis \& Clark College, Portland, OR  97219}

\offprints{lagache@ias.fr}

\date{Received, 27 August 1999; Accepted 4 October 1999}  
          
   \maketitle

\begin{abstract}
We have used the WHAM H$_{\alpha}$ survey and
Leiden/Dwingeloo HI data to decompose the Far-infrared
emission (from 100 to 1000 $\mu$m) at high Galactic latitude
into components associated with the Warm Ionised Medium (WIM)
and the Warm Neutral Medium (WNM). 
This decomposition is possible for the first time
thanks to preliminary WHAM data that cover a significant fraction
of the sky (about 10$\%$). We confirm the first detection
of dust emission from the WIM (Lagache et al. 1999) and show
that the WIM dust temperature and emissivity are very
similar to those in the WNM. The analysis suggests moreover
that about 25$\%$ of the far-IR dust emission at high galactic latitude
is uncorrelated with the HI gas. The decomposition again
reveals a Cosmic Far-Infrared Background (CFIRB)
which is determined for the first time from 100 to
1000 $\mu$m using two independant gas tracers.


\end{abstract}

\section{Introduction}
Two discoveries have recently opened new perspectives on the
understanding dust evolution in the Warm Ionised
Medium (WIM). 
Howk \& Savage (1999) have pointed out, for the first time, the existence
of Al- and Fe-bearing dust grains towards two high-z stars. They have shown
that the degree of grain destruction in the ionised medium through
these two stars is not much higher than in the warm neutral medium. 
In addition, Lagache et al. (1999) have for the first
time detected the infrared dust emission associated with the WIM
based on FIRAS data analysis. This
study reveals a dust emissivity in the WIM
which does not differ much from that of the HI gas.
However, the detection of infrared emission from dust in the WIM is difficult
because one can not easily separate the contribution of the H$^{+}$ gas from that 
of the HI gas. For the first time, the WHAM H$_{\alpha}$ sky survey of our Galaxy
($\delta>-30^o$) offers the unique opportunity to separate the dust emission
associated with the diffuse H$^{+}$ and HI gas. Measuring and characterising 
the dust emission from the WIM could allow to understand the evolution
of the dust grains in the low-density gas. Moreover, dust grain can
play an important role in heating the diffuse ionised gas (through
the photoelectric effect, Reynolds \& Cox, 1992) and can be responsible
at least in part of forbidden line strengths (Sembach et al. 1999; 
Bland-Hawthorn et al. 1997).
\\

Dust emission associated with the HI gas has been extensively
studied using the spatial correlation between the infrared dust emission and the 21 cm HI
emission. Boulanger et al. (1996) using DIRBE and FIRAS data together with the HI
emission as measured by the Leiden/Dwingeloo survey of the northern hemisphere
have found that the dust emission spectrum derived from this correlation 
can be quite well represented by a 
single modified Planck curve characterized by T=17.5~K and  
$\tau/N_{HI}~=~10^{-25}~(\lambda/250\mic)^{-2}$~cm$^2$. 
This emissivity law is very close 
to the one predicted by the Draine $\&$ Lee (1984) dust model. \\

Studies of the far-infrared emission at high Galactic latitude
have also revealed 
a residual component interpreted as the Cosmic Far-InfraRed Background
(Puget et al. 1996; Fixsen et al. 1998; Hauser et al. 1998;
Lagache et al. 1999). This background is produced
by the line of sight accumulation of all extragalactic
objects that are not resolved in the FIRAS or DIRBE beam.
Cosmological implications derived from this background
can be found for example in Dwek et al. (1998) and Gispert et al. 
in preparation.\\

In this paper, we present a decomposition of the far-infrared
sky emission using the HI Leiden/Dwingeloo 
(Hartmann \& Burton, 1994) and the
WHAM H$_{\alpha}$ surveys (Reynolds et al. 1998). This decomposition 
gives new determinations of the dust emission associated
with the diffuse HI and H$^+$ gas as well as the CFIRB.
Our study is limited to very diffuse regions.
Detail comparisons between 
HI, H$_{\alpha}$ and far-infrared filaments
are beyond the scope of this paper and will be adressed
in future works.\\
The paper is organised as follows. First (Sect. 2), we
present the data we have used. Sect. 3 describes
the regions selected for the decomposition as well
as the decomposition method. We then present
the results (Sect. 4). Finally, a summary
is given in Sect. 5.

\section{Data presentation}

\subsection{Far-infrared and HI data}

We have used far-infared data from the DIRBE and FIRAS instruments on board
the COBE satellite. The FIRAS instrument is a polarising Michelson 
interferometer with $7\degr$ resolution and two separate bands 
(from 2.2 to 20~cm$^{-1}$ and from 20 to 96 cm$^{-1}$)
which have a fixed spectral resolution of 0.57~cm$^{-1}$ (Fixsen et al. 1994). 
The study presented here is based on ``pass 4'' data which have been
corrected for the zodiacal light (Kelsall et al. 1998), the Cosmic Microwave Background
and its dipole emission (see the FIRAS explanatory supplement).
DIRBE is a photometer with ten bands covering the range
from 1.25 to 240$~\mic$ with 40 arcmin resolution (Silverberg et al. 1993). 
We only use the last three bands (100, 140 and 240 $\mu$m) for which
the zodiacal emission has been subtracted. These maps are called
``Zodi-Subtracted Mission Average Maps''{\footnote{DIRBE and FIRAS data
are available on the Web site http://www.gsfc.nasa.gov/astro/cobe}}.\\
We decide in our analysis to work at the FIRAS resolution 
(1) to increase the DIRBE map signal to noise ratio and (2) to
combine FIRAS and DIRBE data. Thus, we convolve the DIRBE maps 
with the FIRAS Point Spread Function (PSF). 
The PSF is not precisely known for all wavelengths, so we use the approximation 
suggested by Mather (private communication) 
of a $7\degr$ diameter circle convolved with a line of $3\degr$ length
perpendicular to the ecliptic plane (Mather et al. 1986).\\

The HI data we used are those of the Leiden/Dwingeloo survey which covers the
entire sky down to $\delta=~-30\degr$ with a grid spacing of 30' in both l and
b. The 36' half power beam width 
of the Dwingeloo 25~-~m telescope provides 21~-~cm maps at an angular resolution
which closely matches that of the DIRBE maps. Details of the observations
and correction procedures are given by Hartmann (1994) and by 
Hartmann $\&$ Burton (1997). The 21cm-HI data are convolved 
with the FIRAS PSF. We derive the HI column densities
using 1~K~km~s$^{-1}$=1.82~$10^{18}$~H~cm$^{-2}$ (optically thin emission).\\

\subsection{WHAM data}
The WHAM{\footnote{WHAM informations can be found
on the Web site http://www.astro.wisc.edu/wham}}
Sky Survey provides the first northern hemisphere,
velocity-resolved map of H$_{\alpha}$ from our Galaxy
($\delta>-30^o$). The WHAM instrument is a Fabry-Perot spectrometer
which uses a low noise, high efficiency CCD camera as a 
detector behind a pair of 15 cm diameter Fabry-Perots.
WHAM provides a 12 km s$^{-1}$
velocity resolution with one-degree angular resolution
down to sensitivity limits of 0.2 R (1 R = 10$^6$$/$4$\pi$ ph cm$^{-2}$ s$^{-1}$ 
sr$^{-1}$) in a 30 second exposure.
The one-degree angular resolution nicely matches the DIRBE resolution
and allows direct comparison between these two datasets.
We have used two parts of the WHAM survey centred
on (l,b)=(148$^o$, -20$^o$) and (l,b)=(138$^o$, 36$^o$). 
The two maps (40$^o$$\times$50$^o$ and 60$^o$$\times$20$^o$ respectively)
are called map1 and map2.
For these two regions, preliminary data reduction has been done
including bias subtraction, reflected ring subtraction,
annular-summing and flat-fielding. The geocoronal
H$_{\alpha}$ line is easily removed due to the
velocity separation between this atmospheric emission line
and the Galactic emission line. A first-order polynomial
fits the sky background well within the WHAM 200~km~s$^{-1}$ bandpass.
The typical level of the subtracted background is
about 0.03-0.07~R (km s$^{-1}$)$^{-1}$.
Some spectra in the direction of stars brighter than
the sixth magnitude contain the H$_{\alpha}$ absorption line
from the star. These pointings appear
as isolated 1$^o$ diameter depression in the otherwise
smooth emission at high Galactic latitude.
Intensity calibration is performed using
frequent observations of nebular sources.
All WHAM intensities are tied to the absolute intensity measurement
of the North American Nebula (NGC 7000).
The uncertainty on the H$_{\alpha}$ emission for this preliminary dataset is
about 10$\%$. Details on WHAM can be found in Tufte (1997), Haffner
(1999) and Reynolds et al. (1998).\\

For comparison with COBE data, WHAM data have been first projected on the
DIRBE sixcube. A 5$\times$5 pixel median filtering has been applied
in order to remove the defects induced by stellar H$_{\alpha}$ absorption
lines from several lines of sight. These data
are then convolved with the FIRAS beam.

\section{Decomposition of the Far-infrared emission}

\subsection{Pixel selection}
The decomposition of the far-infrared emission in very diffuse regions
requires careful selection of pixels on the sky mainly to avoid 
HII regions and IR emission from molecular clouds. 
We apply several criteria:\\

- We remove pixels with $|\beta|<10^o$ to avoid any contamination 
by residual zodiacal emission. \\

- Only high Galactic latitude regions are selected (b lower than -30$^o$
and greater than 25$^o$ for map1 and map2 respectively). \\

- Diffuse parts of the sky are selected following Lagache et al. (1998). To remove 
molecular clouds and HII regions, we use the DIRBE map of the
240~$\mic$ excess with respect to the 60$\mic$ emission:
$\Delta$S=S$_{\nu}$(240)-~4$\times$S$_{\nu}$(60).
This map shows as positive flux regions, the cold component of the dust emission,
and as negative flux regions, regions where the dust is locally
heated by nearby stars (like the HII regions).
Therefore, diffuse emission pixels are selected with $|\Delta S| < 3 \sigma$, 
$\sigma$ being evaluated from the width of the histogram of $\Delta$S
(the distribution of $\Delta$S is gaussian with a ``negative'' and ``positive''
wing corresponding to heated dust regions and cold molecular clouds respectively). 
This criteria is for example important for the map2 where it removes the cold Polaris, 
Ursa major and Camelopardalis clouds. It removes 6.4$\%$ of the original
high ecliptic and galactic latitude maps. \\  

- We keep all regions for which the H${\alpha}$ emission is between
0.2 and 2~R. The lower value (0.2~R) ensures that we are well
above the noise.
The threshold of 2 R is arbitrary but ensures that the remaining pixels
are not coming from bright enhanced H${\alpha}$ regions caused by
an increase of the electron density.\\

- Pixels with high HI column densities (N(HI)~$>$~6~10$^{20}$ H~cm$^{-2}$)
are removed. We use this HI threshold rather than that derived
in Boulanger et al. (1996) and Lagache et al. (1999) since it increases
the statistics by 30$\%$. With such a criterium, the contribution
of the cold dust emission is lower than 3$\%$ in the far-infrared
bands. \\

We keep 122 FIRAS pixels, corresponding to 2$\%$ of the sky.

\subsection{Conversion of the WHAM H$_\alpha$ intensity into N(H$^+$)}
IR emission that can be correlated with the ionised gas
cannot be simply related to this gas phase as for the HI
component. Emission from HI gas is proportional to N(HI), whereas
WIM emission is derived from the emission
measure which is proportional to $n_e^2$ (where $n_e$ is the 
WIM electron density). Thus, the WIM intensity measured
in Rayleighs, considering a constant electron density along the 
line of sight, is proportional to $N(H^+)n_e$.
For a WIM dust temperature of $T_e$=8000 K,
we compute a conversion factor (from Osterbrock, 1989):
\begin{equation}
\label{conv_1}
I_{\alpha}(R) = 14.5 \times n_e \times N(H^+)_{20cm^{-2}}
\end{equation}
where $N(H^+)_{20cm^{-2}}$ is the $H^+$ column density
normalised at 10$^{20}$ H cm$^{-2}$.
This conversion factor depends weakly on
the temperature. We see from Eq. (\ref{conv_1}) that
it is impossible to disentangle if variations
in I$_{\alpha}$ are due to variations in the electron
density or variations in the H$^+$ column density.\\

For our sample, the mean H$_{\alpha}$ intensity
is equal to 0.69~R with a standard deviation
of 0.37~R. 
For the case where H$_{\alpha}$ intensity
variations are only due to variations in n$_e$ with a constant
H$^+$ column density, we can estimate the infered electron densities. 
For our selected pixels, the minimal and maximal H$_{\alpha}$ intensity is equal to 0.2 R and
1.9 R respectively. This gives a variation in the electron density
by a factor of $\sim$10.
Such a variation is not expected in our
very diffuse high galactic latitude regions 
averaged in the FIRAS 7$^o$ beam. This is supported by the
study of Reynolds (1991) who has derived for the mean
electron density $<n_e>$=0.08 cm$^{-3}$ along high latitude sightlines 
a dispersion of 0.04 cm$^{-3}$. Thus we conclude that most
of the H$_{\alpha}$ intensity variations in the data used for
this study are due to variations in H$^+$ column density
rather than in n$_e$. Using $<n_e>$=0.08 $\pm$ 0.04 cm$^{-3}$,
we compute from Eq. (\ref{conv_1}) the following conversion factor{\footnote{
We can also derive using our H$_{\alpha}$ intensity dataset the 
dispersion for $<n_e>$ by considering that the H$^+$ column density
is constant, and assigning $H_{\alpha}$ intensity variations
to variations in $<n_e>$. We obtain the 
same dispersion of 0.04 cm$^{-3}$.}}:
\begin{equation}
\label{conv_2}
I_{\alpha}(R) = 1.15 \pm 0.6  \times N(H^+)_{20cm^{-2}}
\end{equation}

\subsection{Decomposition method}
Using the conversion factor derived in the previous section, we can describe the 
far-infrared dust emission as a function of the HI
and H$^+$ column density by:
\begin{equation}
\label{main_EQ}
IR= A \times N(HI)_{20cm^{-2}} + B \times N(H^+)_{20cm^{-2}} + C
\end{equation}
where N(HI)$_{20cm^{-2}}$ and N(H$^+$)$_{20cm^{-2}}$ 
are the column densities normalised
at 10$^{20}$ H cm$^{-2}$. The coefficients A, B
and the constant term C are determined simultaneously using regression fits. 
Before performing the infrared regression analysis,
we have correlated the HI with the H$^+$ gas.
The correlation coefficient between these emission lines
is about 38$\%$ for the selected pixels. Such a weak correlation
may come from the cosecant variation of the two components.
Removing the cosecant variations before the analysis results in
a correlation coefficient of zero.\\

At DIRBE wavelengths the fit has been done using a relative weight computed from the
standard deviation maps. The statistic uncertainty on A and B coefficients are very small compared to
real variations of A and B across the sky. However, we do not yet have
enough statistics to test their spatial variations which
will be done using the complete WHAM survey. \\
The A, B and C values obtained using different weights for the fit
are very consistent. Only their statistical uncertainties change significantly.
As we have checked that A, B and C coefficients do not vary with different weights
for DIRBE, the fit has been done using the same weight for each
pixel at the FIRAS wavelengths. \\
The uncertainty in the conversion factor (Eq. (\ref{conv_2})) do not affect the
determination of the A and C coefficients. It only changes the normalisation
of B.\\ 

Before computing the regression fit analysis, we should have removed
the cosecant law variations for each of the components: dust, HI and
H$^+$. However, due to small statistics ($\sim$ 2$\%$ of the sky), 
it is very difficult to accuratly derive the slopes of these relationships.
We have checked using DIRBE data at 100, 140 and 240 $\mu$m
(at the 7$^o$ resolution) that the results derived by removing the
best estimate of the cosecant variations before the fit are in good agreement
with ones obtained without this correction. Thus, we conclude that for
our small dataset, the cosecant law variations do not affect the
result of the decomposition.

\section{Results and discussion}

This section presents the results obtained for A, B and
C and compares them with previous determinations.
The A coefficient is the spectrum of the dust associated with HI gas
for N(HI)=10$^{20}$ H cm$^{-2}$; if detected, B will give the dust emission spectrum
associated with the ionised gas for N(H$^+$)=10$^{20}$ H cm$^{-2}$; and C is
the residual Cosmic Far-InfraRed Background.

\subsection{Spectrum of the dust associated with HI gas}

\begin{figure}  
\epsfxsize=9.cm
\epsfysize=7.cm
\epsfbox{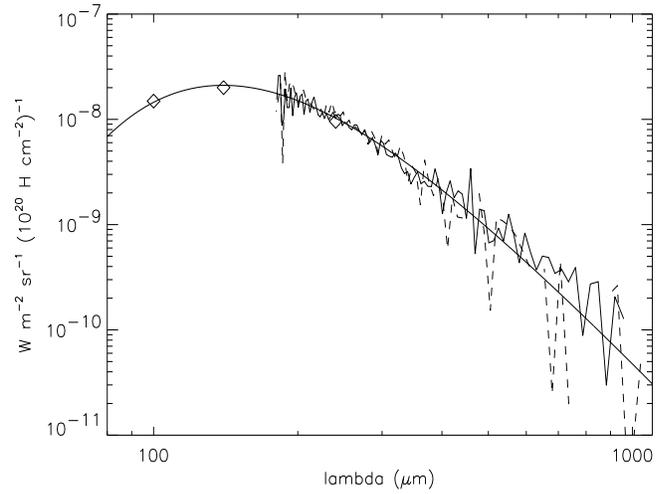}
\caption{\label{hi_spec} Dust emission spectrum associated with HI gas for
N(HI)=10$^{20}$ H cm$^{-2}$. The dashed and continuous lines represent the spectrum derived in
Lagache et al. (1999) and the result of the decomposition (A coefficient) respectively.
Also reported are the DIRBE values at 100, 140 and 240 $\mu$m (diamonds), as well as a
$\nu^2$ modified Planck curve with a temperature of 17.2 K. Errors on
the DIRBE points are smaller than the symbol.}
\end{figure}

The spectrum of the dust associated with HI gas (the A coefficient computed
at each FIRAS wavelength) is shown on Fig. \ref{hi_spec}.
It is limited to $\lambda>$180 $\mu$m due to the very low signal to noise
ratio of FIRAS data at shorter wavelengths. Also reported are A values
obtained at the DIRBE wavelengths (and given in Table \ref{A_B_C_dirbe}). 
We see a very good agreement between
the spectrum obtained from the decomposition and the one derived
by Lagache et al. (1999). To determine the temperature and emissivity
of the dust in HI gas, we fit the FIRAS spectra by a $\nu^2$
modified Planck curve. We obtain a temperature of 17.2 K with an emissivity:
\begin{equation}
\label{HI_emiss}
\tau/N(HI)=8.3 \quad 10^{-26} (\lambda/250\mic)^{-2} \quad cm^2
\end{equation}
This emissivity law is very close 
to the one predicted by the Draine $\&$ Lee (1984) dust model as
was also shown previously by Boulanger et al. (1996).
The relative uncertainty on the emissivity computed
by keeping the modified 17.2 K Planck curve inside the statistical errors is about 10$\%$.

\begin{table}
\begin{center}
\caption{DIRBE dust emission associated with HI (A) and H$^+$ (B) gas for N(HI)=10$^{20}$ H cm$^{-2}$
and N(H$^+$)=10$^{20}$ H cm$^{-2}$ respectively, and CFIRB (C) at 100, 140 and 240 $\mu$m.
Errors on A and B are statistical. On the CFIRB, errors have been estimated using the
width of residual emission ($IR-A \times N(HI) -B \times N(H^+)$) histograms (statistical errors
are negligible).}
\label{A_B_C_dirbe}
\begin{tabular}{|c|c|c|c|} \hline 
$\lambda$ & A MJy/sr/10$^{20}$ & B MJy/sr/10$^{20}$ & C MJy/sr \\ \hline
100 & 0.50$\pm$0.004 & 0.54$\pm$0.016 & 0.78$\pm$0.21 \\ \hline
140 & 0.93$\pm$0.027 & 1.26$\pm$0.103 & 1.13$\pm$0.54 \\ \hline
240 & 0.77$\pm$0.021 & 1.17$\pm$0.078 & 0.88$\pm$0.55 \\ \hline
\end{tabular}\\
\end{center}
\end{table}

\subsection{Spectrum of the dust associated with the WIM}

\begin{figure}  
\epsfxsize=9.cm
\epsfysize=7.cm
\epsfbox{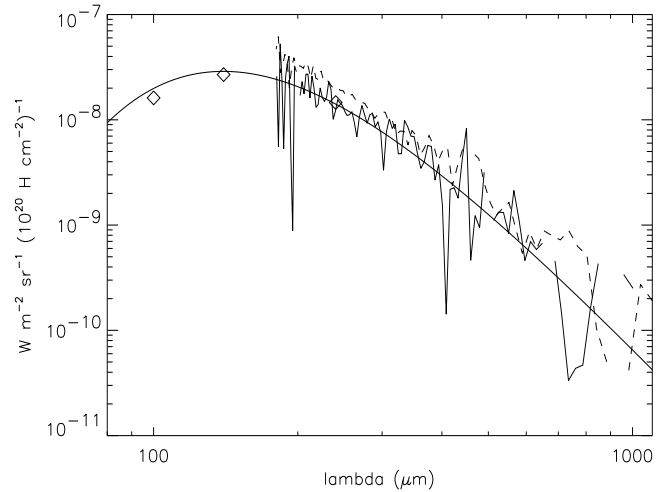}
\caption{\label{hii_spec} Dust emission spectrum associated with the ionised gas for
N(H$^+$)=10$^{20}$ H cm$^{-2}$. The dashed and continuous lines represent the spectrum derived in
Lagache et al. (1999) and the result of the decomposition (B coefficient) respectively.
Also reported are the DIRBE values at 100, 140 and 240 $\mu$m (diamonds), as well as a 
$\nu^2$ modified Planck curve with a temperature of 17.2 K.
Errors on the DIRBE points are smaller than the symbol}
\end{figure}

The B coefficient reveals a spectrum which is clearly non-zero (Fig. \ref{hii_spec}).
This spectrum, although much more noisy, is in very good agreement with the previous 
determination made on a larger fraction of the sky (25.6$\%$, Lagache et al, 
1999). This confirms without ambiguity the first detection of the WIM
dust emission. \\
The level of the spectrum derived here is 1.3 times lower
than the previous one. This factor comes from (1) the use
of the new DIRBE and FIRAS data (for the maps used in Lagache et al. 1999,
the correction of the zodiacal emission was not as accurate as for
the corrected maps delivered by the DIRBE and FIRAS team) and (2)
the difference in the normalisation.
Compared to the previous result, DIRBE data points
(Table \ref{A_B_C_dirbe}) show a decrease of the spectrum at short wavelengths,
suggesting a dust temperature smaller than in the previous study.
This decrease was not seen before in the FIRAS spectrum due
to the very low signal to noise ratio below 180~$\mu$m. 
However, the spectrum given by the B coefficient is
too noisy to determine the dust temperature accuratly. The range of temperature
allowed by the spectrum is between 16 and 18 K.
On Fig. \ref{hii_spec} is displayed, for
direct comparison with the HI gas dust emission a
$\nu^2$ modified Planck curve with a temperature
of 17.2 K. With this temperature, the emissivity
of the dust in the ionised gas is:
\begin{equation}
\tau/N(H^+)=1.1 \quad 10^{-25} (\lambda/250\mic)^{-2} \quad cm^2
\end{equation}
The relative uncertainty on the emissivity computed
by keeping the modified 17.2 K Planck curve inside the statistical errors is
about 10$\%$.
There is no doubt that dust emission associated with the WIM
is detected. However, the uncertainty on the conversion
factor (Eq. (\ref{conv_2})) induces a possible error on the
emissivity value of about 50$\%$.\\

The dust emissivity in the WIM derived for $~<~n_e~>~$=~0.08~cm$^{-3}$ 
is very close to the one derived in the HI gas, which confirms a dust abundance in the WIM
that does not differ much from that of the HI gas.
Such a conclusion is in very good agreement with the recent
work of Howk \& Savage (1999). With their absorption-line
studies of the WIM of the Galaxy, they provide evidence 
for dust in the diffuse ionised gas. Their analysis implies that the processing
of dust grains in the WIM may not be much different than that
in the HI gas. The similar temperature and emissivity of dust
in HI and H$^{+}$ gas support this conclusion. The temperature and emissivity of
the dust in the ionised gas will be more accuratly determined using
the whole WHAM survey.\\

\begin{figure*}
\begin{minipage}{8.cm}
\epsfxsize=9.cm
\epsfysize=7.cm
\hspace{-0.5cm}
\vspace{0.7cm}
\epsfbox{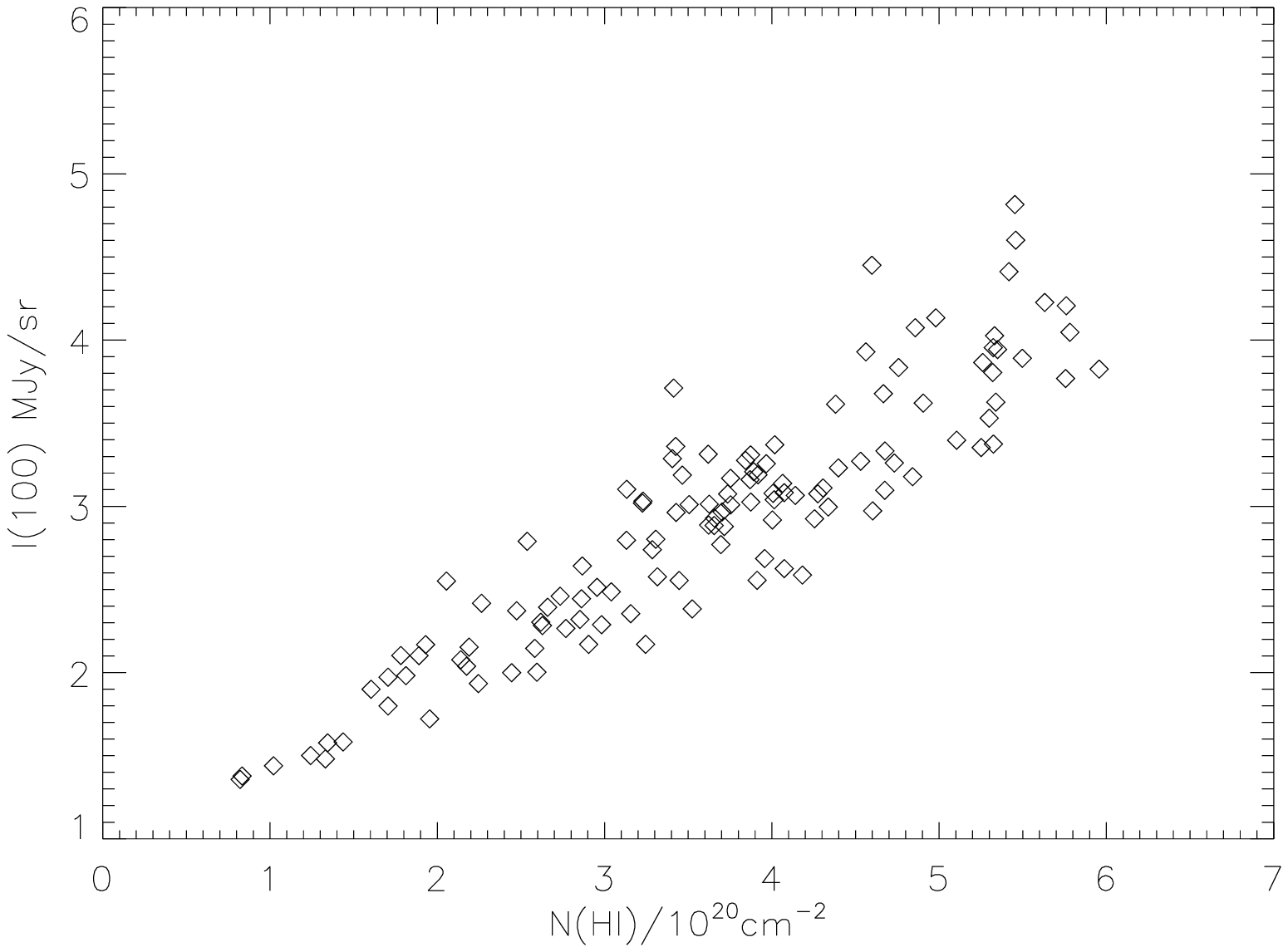}
\end{minipage}
\begin{minipage}{8.cm}
\epsfxsize=9.cm
\epsfysize=7.cm
\hspace{-0.5cm}
\vspace{0.7cm}
\epsfbox{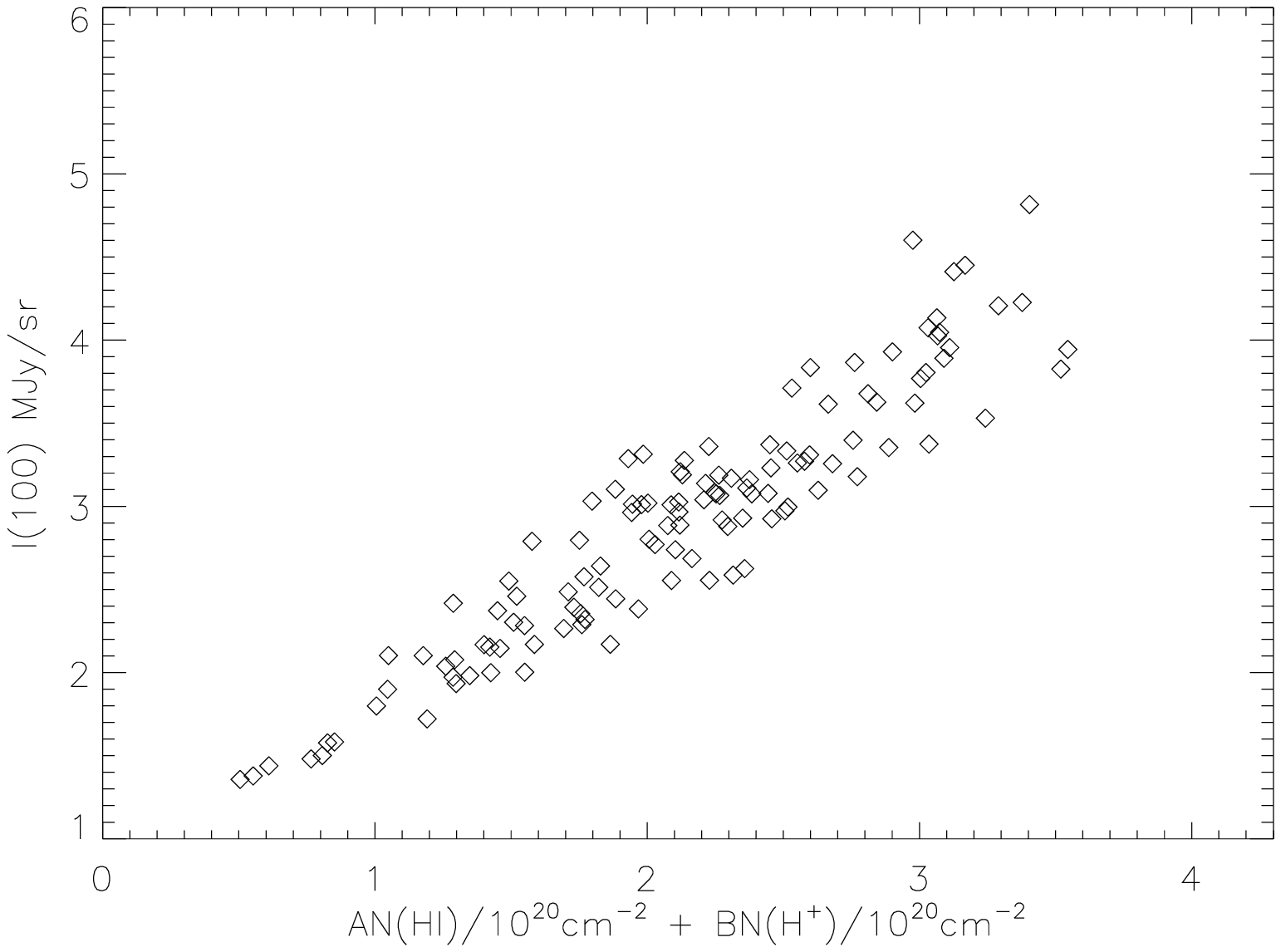}
\end{minipage}\\
\vspace{-1.cm}
\caption{\label{correl_100} a. Correlation between the 100 $\mu$m DIRBE
emission and the HI column density
(normalised at 10$^{20}$ H cm$^{-2}$). b. correlation bewteen the 100 $\mu$m 
emission and the result of the decomposition $A \times N(HI)_{20cm^{-2}} + B \times N(H^+)_{20cm^{-2}}$}
\end{figure*}

These results can be compared with the study of Arendt et al. (1998)
which uses a similar method in order to find the far-infrared
emissivity of the diffuse ionised gas. They conclude that they were unable to
detect any IR emission associated with low density ionised gas at high Galactic latitude.
This non-detection can be easily explained by the very small fraction of the 
sky that was used (the largest region represents about 0.2$\%$ of the sky).\\

Heiles et al. (1999) have done exactly the same decomposition
(Eq. (\ref{main_EQ})) at 100 $\mu$m in the Eridanus superbubble except
that they have used for the infrared emission at 100~$\mu$m the temperature-corrected
DIRBE data from Schlegel et al. (1998). This region is much more complicated
than the diffuse parts of the sky selected here and 
the value of B varies significantly from region to region.
This variation likely reflects the variation of n$_e$
from 0.62 to 1.2 cm$^{-3}$ (Heiles et al. 1999). These 
electron densities are much higher than that in the WIM 
(which is about 0.08 cm$^{-3}$).
The mean B value obtained by Heiles et al. (1999)
is 0.84 MJy/sr/10$^{20}$ H cm$^{-2}$
which is only a factor 1.5 above our determination, which
may be surprising considering the difference in the physical conditions
between our selected parts of the sky  and the Eridanus Superbubble.\\

Fig. \ref{correl_100} shows a comparison beetween the 
100$\mu$m~/~HI and 100$\mu$m~/~[AN(HI)+BN(H$^+$)]
correlations. We clearly see the effect of the dust emission associated
with the ionised gas. The correlation coefficient is higher using
the two gas components rather than using only the HI component
(93.6$\%$ compared to 91.2$\%$). Combined with the value of the
WNM and WIM dust emissivity and assuming that N(H$^+$)/N(HI)$\sim$25$\%$,
this analysis suggests that a significant fraction
(about 20-30 $\%$) of the far-IR dust emission at high galactic latitude
is uncorrelated with the HI gas.\\

\subsection{The Cosmic Far-InfraRed Background}
\begin{figure}  
\epsfxsize=9.cm
\epsfysize=7.cm
\epsfbox{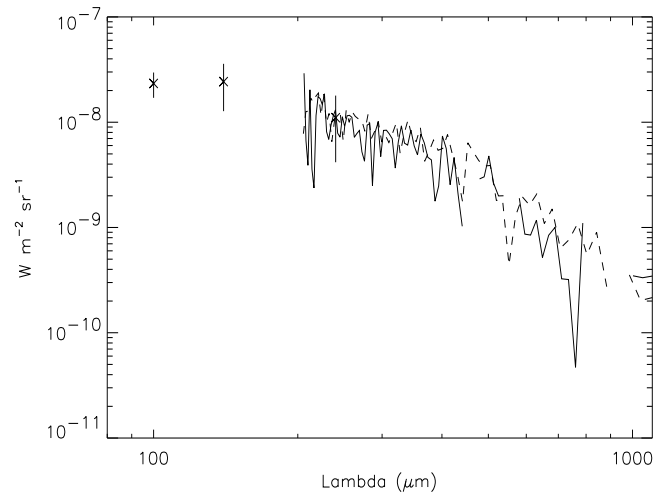}
\caption{\label{CFIRB_spec} CFIRB spectra obtained from the decomposition
of the far-infrared sky (continuous line) and determined on the Lockman
Hole region (dashed line) by Lagache at al. (1999). Also reported
are DIRBE values at 100, 140 and 240 $\mu$m.}
\end{figure}

The C spectrum obtained by the far-infrared emission decomposition
is shown on Fig. \ref{CFIRB_spec} together with the CFIRB determination
of Lagache et al. (1999) in the Lockman Hole region. We see a very
good agreement between the two spectra. These determinations are also
in good agreement with the Fixsen et al. (1998) one.
This confirmation of the level and shape of the CFIRB supports the
validity of the decomposition using the two independant
datasets since the constant term in Eq. (\ref{main_EQ})
is very dependant on the determination of A and B.\\

At 140 and 240 $\mu$m, the values obtained for the CFIRB
are 1.13$\pm$0.54 MJy/sr and 0.88$\pm$0.55 MJy/sr respectively 
(Table \ref{A_B_C_dirbe}). For each selected pixel, we compute
the residual emission, R = IR - A$\times$N(HI) - B$\times$N(H$^+$).
Uncertainties of the CFIRB have been derived from the width of the histogram
of R (statistical uncertainties derived from the regression analysis are negligible). 
The obtained CFIRB values, although much more noisy
(due to the small fraction of the sky used), are in very
good agreement with Hauser et al. (1998) determinations
and the Lagache et al. (1999) one at 240 $\mu$m.
At 140~$\mu$m, the CFIRB value of Lagache et al. (1999) is smaller 
than that derived here since the assumed WIM dust spectrum
were overestimated (the WIM dust spectrum was
very noisy below 200~$\mu$m and the estimated dust
temperature was too high).\\

At 100~$\mu$m, DIRBE data are less noisy than at 140 and
240~$\mu$m (the detector at 100 $\mu$m is a photoconductor
rather than bolometers at 140 and 240~$\mu$m).
Assuming an accurate subtraction of the zodiacal emission
at this wavelength, our decomposition gives the CFIRB at 100~$\mu$m: 
I$_{CFIRB}$(100)= 0.78$\pm$0.21
MJy/sr. This value can be compared to the non-isotropic residual
emission found by Hauser et al. (1998). The average on three regions
of the residual emission, equal
to 0.73$\pm$0.20 MJy/sr, is in very good agreement with our
determination.

\section{Summary}
We have performed a decomposition of the far-infrared sky
at high Galactic latitude using two independant gas
tracers: the Leiden/Dwingeloo HI and WHAM H${\alpha}$
surveys of the northern hemisphere of our Galaxy.
Main results of this analysis are:\\

1) The dust emission associated with the HI gas is very well represented
by a $\nu^2$ modified Planck curve with a temperature of 17.2 K
and an emissivity:\\
$\tau/N(HI)=8.3 \quad 10^{-26} (\lambda/250\mic)^{-2} \quad cm^2$\\
This emissivity is in very good agreement with previous determinations (Lagache et al. 1999;
Boulanger et al. 1996) and also with the Draine \& Lee (1984) dust model. \\

2) We confirm the first detection of dust emission from the WIM.
Assuming a temperature of 17.2 K, the WIM dust emissivity is:\\
$\tau/N(HI)=1.1 \quad 10^{-25} (\lambda/250\mic)^{-2} \quad cm^2$\\
suggesting a dust abundance in the WIM comparable to that in the
HI gas.\\

3) The CFIRB obtained from this decomposition is in very good
agreement with the previous determinations (Fixsen et al. 1998; 
Hauser et al. 1998; Lagache et al. 1999). This is the first 
time that two independent gas tracers for the HI and the
H$^+$ have been used to determine the background at 100 
$\mu$m: I$_{CFIRB}$=0.78$\pm$0.21 MJy/sr.\\

The complete WHAM survey will allow an extension of this decomposition
to the whole northern sky.
In particular, variation of the dust emission spectrum
associated with different ionisation environments will give
unprecedent information on the dust processing in the ionised
gas of our Galaxy.\\

Acknowledgements:\\
G.L. is very grateful to R. Reynolds, M. Haffner and S. Tufte
for hospitality extended during the stay 
at the Wisconsin University and for providing early WHAM survey data.
Many thanks to David Leisawitz for ``pass 4'' FIRAS data and to A. Abergel,
F. Boulanger and J.L. Puget for their carefull reading of the manuscript.
WHAM is supported by a grant from the National Science Foundation.\\



\begin{thebibliography}{}
\bibitem[1998]{Arendt} Arendt R.G., Odegard N., Weiland J.L., et al. 1998, ApJ 508, 74
\bibitem[1996]{Boulanger96} Boulanger F., Abergel A., Bernard J.P., et al. 1996, A\&A 312, 256 
\bibitem[1999]{Bland} Bland-Hawthorn J., Freeman K.C., Quinn P.J., 1997, ApJ 490, 143
\bibitem[1984]{draine84} Draine B.T., Lee H.M., 1984, ApJ 285, 89
\bibitem[1998]{Dwek} Dwek E., Arendt R.G., Hauser M.G., et al., 1998, ApJ 508, 106 
\bibitem[1994]{Fixsen94} Fixsen D.J., Cheng E.S, Cottingham D.A., et al. 1994, ApJ 420, 457
\bibitem[1998]{Fixsen98} Fixsen D.J., Dwek E., Mather J.C., et al. 1998, ApJ 508, 123
\bibitem[1998]{haffner} Haffner L.M., Reynolds R.J., Tufte S.L., 1998, ApJ 501, L83
\bibitem[1999]{haff} Haffner L.M., 1999, Ph.D. Thesis, University of Wisconsin
\bibitem[1994]{Hartm94} Hartmann D., 1994, Ph.D. Thesis, University of Leiden
\bibitem[1997]{Hartmann} Hartmann D., Burton W.B., Atlas of Galactic neutral
Hydrogen, Cambridge University Press, 1997
\bibitem[1998]{Hauser98} Hauser M.G., Arendt R.G., Kelsall T., et al., 1998, ApJ 508, 25
\bibitem[1999]{Heiles} Heiles C., Haffner L.M., Reynolds R.J., 1999, 
in New Perspectives on the Interstellar Medium, ASP Conference
Series, Vol. 168, Eds A.R. Taylor, T.L. Landecker, G. Joncas
\bibitem[1999]{ho} Howk J.C., Savage B.D., 1999, ApJ 517, 746
\bibitem[1998]{Kelsall} Kelsall T., Weiland J.L., Franz B.A., et al., 1998, ApJ 508, 44
\bibitem[1998]{lag} Lagache G., Abergel A., Boulanger F., Puget J.L., 1998,
A\&A 333, 709
\bibitem[1999]{laga} Lagache G., Abergel A., Boulanger F., et al., 1999, A\&A 344, 322
\bibitem[1986]{Mather86} Mather J.C., et al. 1986, App Opt 25, 16
\bibitem[1989]{oster} Osterbrock D.E, Astrophysics of Gaseous Nebulae
and Active Galactic Nuclei, University science book, 1989
\bibitem[1996]{Puget96} Puget J.L., Abergel A., Bernard J.P., et al. 1996, A\&A 308, L5 
\bibitem[1991]{Rey91} Reynolds R.J., 1991, ApJ 372, L17
\bibitem[1991]{Rey91} Reynolds R.J., Cox D.P., 1992, ApJ 400, L33
\bibitem[1998]{Rey} Reynolds R.J., Tufte S.L., Haffner L.M., et al., 1998, PASA 15, 14
\bibitem[1999]{Sem} Sembach K.R., Howk J.C., Ryans R.S.I., Keenan F.P., ApJ in press
\bibitem[1999]{Schlegel} Schlegel D.J., Finkbeiner D.P., Davis, M. 1998, ApJ 500, 525
\bibitem[1993]{Silverberg93} Silverberg R.F., et al. 1993, in SPIE 
Conference Proc. 2019 on Infrared Spaceborne Remote Sensing, San 
Diego
\bibitem[1997]{Tufte} Tufte S.L, 1997, Ph.D. thesis, University of Wisconsin
\end{thebibliography}
\end{document}